\documentclass[doublecol]{epl2}
\usepackage{graphicx}
\usepackage{amsmath}
\usepackage{amssymb}
\usepackage{bm,hyperref}

\title{Exotic superfluidity in spin-orbit coupled Bose-Einstein condensates}

\author{Qizhong Zhu\inst{1} \and Chuanwei Zhang\inst{2} \and Biao Wu\inst{1}}

\institute{
  \inst{1} International Center for Quantum Materials, Peking University, 100871, Beijing, China\\
  \inst{2} Department of Physics and Astronomy, Washington State University, Pullman, Washington, 99164 USA
}

\abstract{
We study the superfluidity of a spin-orbit coupled Bose-Einstein condensate (BEC)
by computing its Bogoliubov excitations, which are found to consist of two branches:
one is gapless and phonon-like at long wavelength; the other is typically gapped. These
excitations imply a superfluidity that has two new features: ({\it i})  due to the absence
of the Galilean invariance, one can no longer define the critical velocity of superfluidity
independent of the reference frame; ({\it ii}) the superfluidity depends not only
on whether the speed of the BEC exceeds a critical value, but also on
{\it cross helicity} that is defined as the direction of the cross product of the spin and
the kinetic momentum  of the BEC.}

\begin{document}

\maketitle

Superfluidity was first discovered in the 1930s, and has
fascinated physicists ever since. This interesting phenomenon was explained
by Landau \cite{landau}, whose theory has been very successful in explaining
many important properties of superfluids. However, Landau's theory of
superfluidity may be facing challenges brought by the recent experimental
realization of artificial gauge fields for ultracold bosonic atoms \cite%
{lin1,lin2,fu,pan}. When the artificial gauge field is non-Abelian
\cite{ruseckas,dalibard,zhang1}, it is effectively spin-orbit coupling (SOC).
SOC has played a crucial role in many exotic phenomena such as topological
insulators \cite{ti}. However, in superfluids, the SOC is generally absent
and its effects have remained largely unexplored. Note that this issue is
not limited to ultracold atoms since the Bose-Einstein condensation with SOC
also exists for excitons in semiconductors \cite{yao,shem}.

There have been some theoretical works, where many interesting properties of
spin-orbit coupled BECs are explored \cite%
{shem,zhai1,ho,victor,zhang,yip,zhai2,baym,pu,you,zhangqi,santos,fleischhauer}. For instance, it was
shown in Ref. \cite{shem} that SOC can lead to unconventional Bose-Einstein
condensation with the breaking of time-reversal symmetry. Later a stripe
phase that breaks rotational symmetry was found \cite{zhai1,ho}. In this
Letter we concentrate on the superfluidity of the spin-orbit coupled BEC.

To put our study into perspective, we briefly review Landau's theory for a
superfluid without SOC. Consider such a superfluid flowing in a tube. With
the Galilean transformation, Landau found that the excitation of this
flowing superfluid is related to the excitation of a motionless superfluid
as \cite{landau}
\begin{equation}
\varepsilon_{v}({\mathbf{p}})=\varepsilon_{0}({\mathbf{p}})+{\mathbf{p}}%
\cdot {\mathbf{v}}\,,  \label{landau}
\end{equation}
where $\mathbf{p}$ is the momentum of the excitation. For phonon excitation $%
\varepsilon_{0}({\mathbf{p}})=c|\mathbf{p}|$, the excitation $%
\varepsilon_{v}({\mathbf{p}})$ can be negative only when $|{\mathbf{v}}|>c$.
Therefore, the speed of sound $c$ is the critical velocity beyond which the
flowing superfluid loses its superfluidity and suffers viscosity. We switch
to a different reference frame, where the superfluid is at rest while the
tube is moving. It is apparent to many that these two reference frames are
equivalent so that the superfluid will be dragged along only when the tube
speed exceeds the speed of sound $c$. However, this equivalence is based on
that the superfluid is invariant under the Galilean transformation. As SOC
breaks the Galilean invariance of the system \cite{messiah}, we find that
these two reference frames are no longer equivalent as shown in Fig. \ref%
{soc}: the critical speed for (a) is different from the one for (b). For
easy reference, the critical speed for (a) is hereafter called the critical
flowing speed and the one for (b) the critical dragging speed.

\begin{figure}[!h]
\includegraphics[width=8cm]{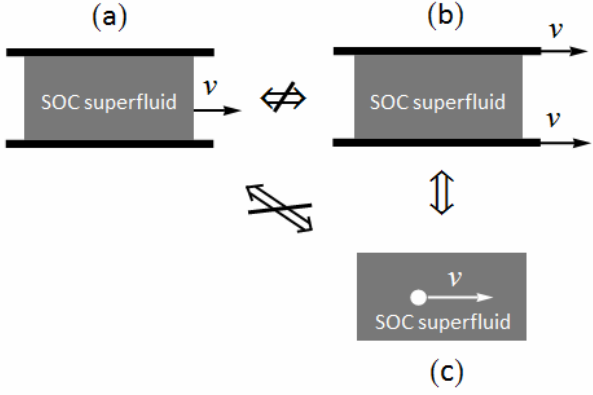}\newline
\caption{(a) A superfluid with spin orbit-coupling moves while the tube is
at rest. (b) The superfluid is dragged by a tube moving at the speed of $v$.
(c) An impurity moves at $v$ in the SOC superfluid. The reference frame is
the lab. The two-way arrow indicates the equivalence between different
scenarios and the arrow with a bar indicates the non-equivalence.}
\label{soc}
\end{figure}

Our study of the superfluidity for a BEC with SOC is based on the
computation of the elementary excitations using the Bogoliubov equation. We
calculate how the elementary excitations change with the flow speed and
manage to derive from these excitations the critical speeds for the two
different scenarios shown in Fig. \ref{soc}(a,b). We find that there are two
branches of elementary excitations for a BEC with SOC: the lower branch is
phonon-like at long wavelengths and the upper branch is generally gapped.
Careful analysis of these excitations indicates that a moving BEC with the
SOC (Fig. \ref{soc}(a)) has non-zero critical velocities for superfluidity
while the critical speed is zero for a BEC being dragged by a moving
tube(Fig. \ref{soc}(b)). This shows that the definition of critical velocity
depends on the reference frame for a superfluid with SOC and, probably, for
any superfluid that has no Galilean invariance.

In addition, we find that the properties of a flow of BEC with SOC are also
related to its spin direction. We characterize this spin direction with
cross helicity, which is defined as the cross product of the spin and the
kinetic momentum of the flow. A BEC flow with Rashba SOC is always unstable
if its cross helicity is negative.

\textit{Model}:~We consider a BEC with pseudospin $1/2$ and the Rashba SOC.
The dynamics of the system can be described by the Hamiltonian \cite%
{zhai1,merkl,larson,zhang}
\begin{eqnarray}
\mathcal{H} &=&\int d\mathbf{r}\left\{ \sum_{\sigma =1,2}\Psi _{\sigma
}^{\ast }\left( -\frac{\hbar ^{2}\nabla ^{2}}{2m}+V(\mathbf{r})\right) \Psi
_{\sigma }\right.   \notag \\
&&\left. +\gamma \left[ \Psi _{1}^{\ast }(i\hat{p}_{x}+\hat{p}_{y})\Psi
_{2}+\Psi _{2}^{\ast }(-i\hat{p}_{x}+\hat{p}_{y})\Psi _{1}\right] \right.
\notag \\
&&\left. +\frac{C_{1}}{2}\left( |\Psi _{1}|^{4}+|\Psi _{2}|^{4}\right)
+C_{2}|\Psi _{1}|^{2}|\Psi _{2}|^{2}\right\} \,,  \label{ham}
\end{eqnarray}%
where $\gamma $ is the SOC constant, $C_{1}$ and $C_{2}$ are interaction
strengths between the same and different pseudospin states, respectively.

We focus on the homogeneous case $V(\mathbf{r})=0$ despite that the BEC
usually resides in a harmonic trap in experiments. The primary reason is
that the superfluidity can be discussed more clearly in the homogeneous
case, and be compared directly with the conventional superfluidity of a
spinless bosonic system. In addition, the results in the homogeneous case
can be adopted to understand the superfluidity in more complicated
situations with the local density approximation. We also limit ourselves to
the case $C_{1}>C_{2}$, where the system is stable against phase separation
\cite{zhai1,zhang}. In the following discussion, for simplicity, we set $%
\hbar =m=1$ and ignore the non-essential $z$ direction, treating the system
as two-dimensional. This does not impair the validity of our model, since in the
following we assume the BEC moves in the $y$ direction, and the
critical velocity is found to be not influenced by the excitation in the $z$ direction.

\begin{figure*}[!htb]
\center{\includegraphics[width=14cm]{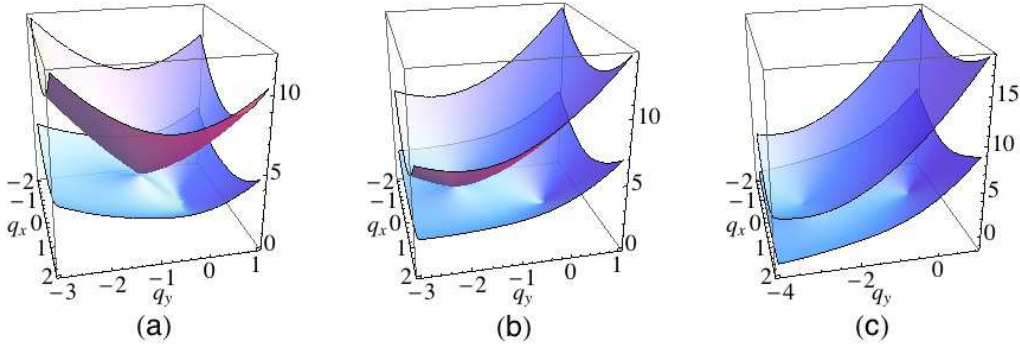}}
\caption{(color online)Elementary excitations of a BEC flow with the SOC.
(a) $k=1$; (b) $k=2.5$; (c) $k=4$. $C_1=10$, $C_2=4$, $\protect\gamma=1$. }
\label{3D}
\end{figure*}

The Gross-Pitaevskii equation obtained from the Hamiltonian (\ref{ham}) has
plane wave solutions
\begin{equation}
\phi_{{\mathbf{k}}}=\left(%
\begin{array}{c}
\Psi_1 \\
\Psi_2 %
\end{array}%
\right)
=\frac{1}{\sqrt{2}}\left(%
\begin{array}{c}
e^{i\theta_{\mathbf{k}}} \\
- 1%
\end{array}%
\right)e^{i\mathbf{k}\cdot\mathbf{r}-i\mu(\mathbf{k}) t}\,,  \label{pwave}
\end{equation}
where $\tan\theta_{\mathbf{k}}=k_{x}/k_{y}$, $\mu(\mathbf{k})=|\mathbf{k}%
|^2/2-\gamma|\mathbf{k}|+(C_1+C_2)/2$. The solution $\phi_{{\mathbf{k}}}$ is
the ground state of the system when $|\mathbf{k}|=\gamma$. There is another
set of plane wave solutions, which have higher energies and are not relevant
to our discussion.

The plane wave solution $\phi _{{\mathbf{k}}}$ represents a BEC flow with
the velocity $\mathbf{v}=\mathbf{k}-\gamma \hat{\mathbf{k}}$($\hat{\mathbf{k}%
}=\mathbf{k}/|\mathbf{k}|$). This velocity is essentially the kinetic
momentum of the BEC, which is different from the conjugate momentum $\mathbf{%
k}$ due to the presence of SOC. For example, the ground state has a non-zero
conjugate momentum $\mathbf{k}$, but its kinetic momentum (or velocity) is
zero. Besides its velocity, the BEC flow described by $\phi _{{\mathbf{k}}}$
has another feature, the direction of the spin, which has to be included for
a complete description of the flow. For example, the flow at $\mathbf{k}%
=3\gamma /2\hat{y}$ has the same velocity $\gamma /2\hat{y}$ as the flow at $%
\mathbf{k}=-\gamma /2\hat{y}$. However, they have different spin directions.
With the Rashba type SOC, the spin direction of the eigenstate is always
perpendicular to the velocity and has only two choices. As a result, it is
sufficient and also very convenient to use $w=\mathrm{sign}[(\hat{\mathbf{v}}%
\times \hat{\sigma})\cdot \hat{z}]$ to denote the spin direction. This
variable $w$, which is either $1$ or $-1$, is the cross helicity mentioned
in the introduction. Note that the usual helicity in literature is defined
as the inner product of the spin and the momentum.

\textit{Critical velocities}:~We study first the scenario depicted in Fig. %
\ref{soc}(a), where the BEC flows with a given velocity. Since the system is
not invariant under the Galilean transformation, we cannot use Eq. (\ref%
{landau}) to find the excitations for the flowing BEC from the excitation of
a stationary BEC. We have to compute the excitations directly. This can be
done by computing the elementary excitations of the state $\phi _{{\mathbf{k}%
}}$ with the Bogoliubov equation for different values of $\mathbf{k}$.

Without loss of generality, we choose $\mathbf{k}=k\hat{y}$ with $k>0$.
Following the standard procedure of linearizing the Gross-Pitaevskii
equation \cite{wu1,wu2}, we have the following Bogoliubov equation%
\begin{equation}
\mathcal{L}\left(
\begin{array}{c}
u_{1} \\
u_{2} \\
v_{1} \\
v_{2}%
\end{array}%
\right) =\varepsilon \left(
\begin{array}{c}
u_{1} \\
u_{2} \\
v_{1} \\
v_{2}%
\end{array}%
\right) ,
\end{equation}%
where
\begin{equation}
\mathcal{L}=\left(
\begin{array}{cccc}
H_{k}^{+} & b_{12} & -\frac{1}{2}C_{1} & -\frac{1}{2}%
C_{2} \\
 b_{21} & H_{k}^{+} & -\frac{1}{2}C_{2} & -\frac{1}{2}%
C_{1} \\
\frac{1}{2}C_{1} & \frac{1}{2}C_{2} & H_{k}^{-} &  b_{34} \\
\frac{1}{2}C_{2} & \frac{1}{2}C_{1} & b_{43} &
H_{k}^{-}%
\end{array}%
\right)
\end{equation}%
with $H_{k}^{\pm }=\pm \frac{q_{x}^{2}+(q_{y}\pm k)^{2}}{2}\pm A$, $A=\frac{%
C_{1}}{2}-\frac{k^{2}}{2}+\gamma k$, $b_{12}=-\gamma(iq_{x}+q_{y}+k)+\frac{C_{2}}{2}$, $%
b_{21}=\gamma(iq_{x}-q_{y}-k)+\frac{C_{2}}{2}$,
 $b_{34}=\gamma(iq_{x}-q_{y}+k)-\frac{C_{2}}{2}$, and $b_{43}=-\gamma(iq_{x}+q_{y}-k)-\frac{C_{2}}{2}$.
 As usual, there are two groups of eigenvalues and only the ones whose
corresponding eigenvectors satisfy $|u_{i}|^{2}-|v_{i}|^{2}=1$ ($i=1,2$) are
physical.

For comparison, we consider the case without SOC. This is to put $\gamma =0$
and reduce our system to a two-component BEC system, which is well studied
in literature \cite{pu2}. For this case, $\mathcal{L}$ can be diagonalized
analytically and there are two branches of excitations
\begin{equation}
\varepsilon _{\pm }({\mathbf{q}})=q_{y}k+\sqrt{\frac{C_{1}\pm C_{2}}{2}{%
\mathbf{q}}^{2}+\frac{{\mathbf{q}}^{4}}{4}}\,.
\end{equation}%
These results show that the system at the ground state ($k=0$) has two
different speeds of sound, $\sqrt{(C_{1}+C_{2})/2}$ and $\sqrt{%
(C_{1}-C_{2})/2}$. Since the excitation $\varepsilon _{-}$ becomes negative
only when $k>\sqrt{(C_{1}-C_{2})/2}$, the critical velocity in this case is $%
\sqrt{(C_{1}-C_{2})/2}$. When $C_2=0$, these two branches of excitations
merge into one and the critical velocity is $\sqrt{C_{1}/2}$, which is well
known and was confirmed in a BEC experiment \cite{raman}.

\begin{figure}[!htb]
\includegraphics[width=3.4in]{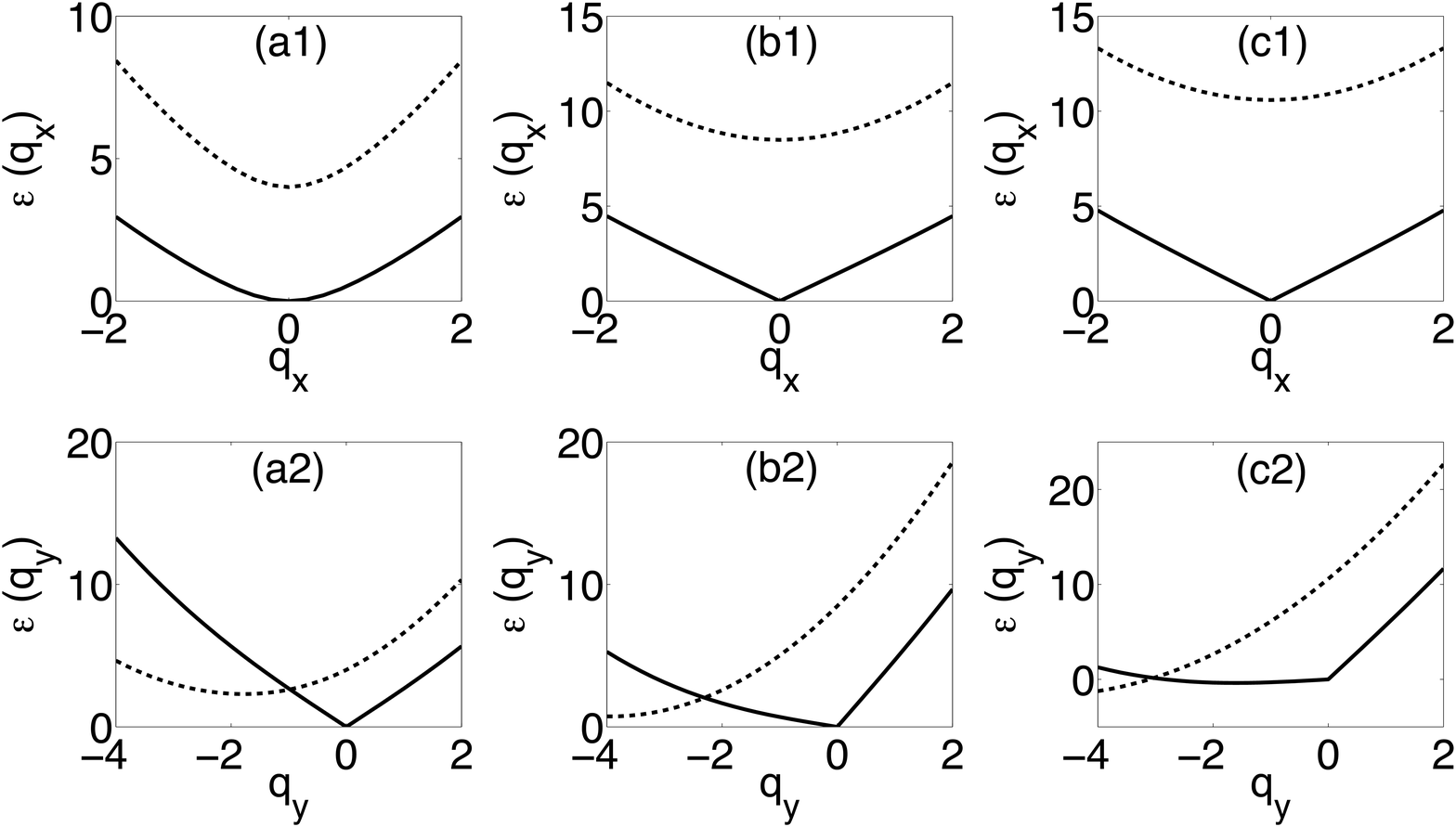}\newline
\caption{Excitations along the $x$ direction (the first row) and $y$
direction (the second row) at different values of $k$. (a1, a2) $k=1$; (b1,
b2) $k=3$; (c1, c2) $k=4$. $C_1=10$, $C_2=4$, $\protect\gamma=1$.}
\label{xy}
\end{figure}

In general, there are no simple analytical results. We have numerically
diagonalized $\mathcal{L}$ to obtain the elementary excitations. We find
that part of the excitations are imaginary for BEC flows with $|{\mathbf{k}}%
|<\gamma$. This means that all the flows with $|{\mathbf{k}}|<\gamma$ are
dynamically unstable and therefore do not have superfluidity. For other
flows with $k\ge \gamma$, the excitations are always real and they are
plotted in Fig. \ref{3D}. One immediately notices that the excitations have
two branches, which contact each other at a single point. More closer
examination shows that the upper branch is gapped in most of the cases while
the lower branch has phonon-like spectrum at large wavelength. These
features are more apparent in Fig. \ref{xy}, where only the excitations
along the $x$ direction and $y$ direction are plotted.

In Fig. \ref{3D}(c) and Fig. \ref{xy}(c2), we notice that part of the
excitations in the upper branch are negative, indicating that the underlying
BEC flow is thermodynamically unstable and has no superfluidity. In fact,
our numerical computation shows that there exists a critical value $k_{c}$:
when $k>k_{c}$ either part of the upper branch of excitations or part of the
lower branch or both become negative. This means that the flows described by
the plane wave solution $\phi _{{\mathbf{k}},-}$ with $|{\mathbf{k}}|>k_{c}$
suffer the Landau instability and have no superfluidity. Combined with the
fact that the flows with $|{\mathbf{k}}|<\gamma $ are dynamically unstable,
we can conclude that only the flows with $\gamma \leq \mathbf{|k|}\leq k_{c}$
have superfluidity. Physically, these super-flows have speeds smaller than $%
v_{c}=k_{c}-\gamma $ and cross helicity $w=1$. We have plotted how the
critical flowing velocities vary with the SOC parameter $\gamma$ in Fig. \ref%
{vc}.

\begin{figure}[!t]
\includegraphics[width=3.4in]{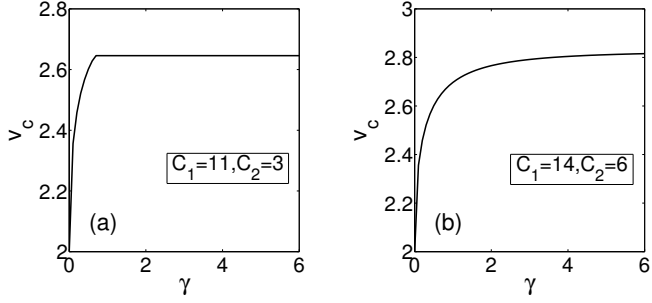}\newline
\caption{Critical flowing velocity $v_c$ (solid line) of a BEC as a function
of the SOC parameter $\protect\gamma$. (a) $C_1>3C_2$; (b) $C_1<3C_2$.}
\label{vc}
\end{figure}

We find two different asymptotic behaviors for the critical velocity when $%
\gamma$ gets large. As shown in Fig. \ref{vc}, the critical flowing velocity
becomes a constant, $v_c=\sqrt{(C_{1}+C_{2})/2}$, beyond a threshold value
in the parameter regime $C_1>3C_2$ while it approaches an asymptotic value $%
\sqrt{C_{1}-C_{2}}$ in the regime $C_1<3C_2$.

We turn to another reference frame illustrated in Fig. \ref{soc}(b), where
the BEC can be viewed as being dragged by a moving tube. To simplify the
discussion, we replace the moving tube with a macroscopic impurity moving
inside the BEC as shown in Fig. \ref{soc}(c). Correspondingly, the question
``whether the BEC will be dragged along by the moving tube?" is replaced by
an equivalent question ``whether the impurity will experience any
viscosity?". Suppose the moving impurity generates an excitation in the BEC.
According to the conservations of both momentum and energy, we should have
\begin{eqnarray}  \label{con1}
m_0{\mathbf{v}}_i&=&m_0{\mathbf{v}}_f+{\mathbf{q}}\,, \\
\frac{m_0{\mathbf{v}}_i^2}{2}&=&\frac{m_0{\mathbf{v}}_f^2}{2}+\varepsilon_0({%
\mathbf{q}})\,,  \label{con2}
\end{eqnarray}
where $m_0$ is the mass of the impurity, ${\mathbf{v}}_i$ and ${\mathbf{v}}_f
$ are the initial and final velocities of the impurity, respectively, and $%
\varepsilon_0({\mathbf{q}})$ is the excitation of the BEC at $k=\gamma$. If
the excitations were purely phonons, i.e., $\varepsilon_0({\mathbf{q}})=c|{%
\mathbf{q}}|$, these two conservations would not be satisfied simultaneously
when $v\approx|{\mathbf{v}}_i|\approx|{\mathbf{v}}_f|<c$. This means that
the impurity could not generate phonons in the superfluid and would not
experience any viscosity when its speed were smaller than the sound speed.
This is in fact nothing but the Cerenkov radiation \cite{key2,key3}, where a
charged particle radiates only when its speed exceeds the speed of light in
the medium. For our BEC system, the elementary excitations $\varepsilon_0({%
\mathbf{q}})$ are not purely phonons. In this case, the critical dragging
velocity derived from Eqs. (\ref{con1},\ref{con2}) is given by
\begin{equation}
v_c=\left|\frac{\varepsilon_0({\mathbf{q}})}{|{\mathbf{q}}|}\right|_{min}\,.
\label{cerenkov}
\end{equation}

For the special case $\gamma=0$, we have
\begin{equation}
\varepsilon _{0}^{\pm }({\mathbf{q}})=\sqrt{\frac{C_{1}\pm C_{2}}{2}{\mathbf{q%
}}^{2}+\frac{{\mathbf{q}}^{4}}{4}}\,.
\end{equation}
From Eq. (\ref{cerenkov}), we obtain the critical dragging speed $\sqrt{%
\frac{C_{1}- C_{2}}{2}}$, which is the same as the critical flowing speed.

When $\gamma\neq 0$, The excitations $\varepsilon_0({\mathbf{q}})$ also
share two branches. Along the $x$ direction, these two branches are
\begin{equation}
\varepsilon _{0}^{\pm }(q_{x})=\sqrt{s_{1}+s_{2}q_{x}^{2}+\frac{q_{x}^{4}}{4}%
\pm \sqrt{t_{1}+t_{2}q_{x}^{2}+t_{3}q_{x}^{4}+\gamma ^{2}q_{x}^{6}}},
\end{equation}%
where $s_{1}=2\gamma ^{4}+\gamma ^{2}\left( C_{1}-C_{2}\right) $, $%
s_{2}=2\gamma ^{2}+\frac{1}{2}C_{1}$, $t_{1}=s_{1}^{2}$, $t_{2}=2s_{1}s_{2}$%
, and $t_{3}=2s_{1}+\left( \gamma ^{2}+C_{2}/2\right) ^{2}$.
Along the $y$ direction, the excitations of the ground state are
\begin{eqnarray}
\varepsilon _0^{-}(q_{y}) &=&\sqrt{\frac{C_{1}+C_{2}}{2}q_{y}^{2}+\frac{q_{y}^{4}%
}{4}}\,, \\
\varepsilon _0^{+}(q_{y}) &=&2\gamma q_{y}+\sqrt{2s_{1}+\left( s_{2}-\frac{C_{2}}{%
2}\right) q_{y}^{2}+\frac{q_{y}^{4}}{4}}\,.
\end{eqnarray}%
When $\gamma >0$, the upper branch $\varepsilon _{0}^{+}(q_{x})$ is always parabolic at
small $q_{x} $ with a gap $\sqrt{2s_{1}}$. When expanded to the second order
of $q_{x}$, the lower branch has the following form
\begin{equation}
\varepsilon _{0}^{-}(q_{x})\approx q_{x}^{2}\sqrt{\frac{C_{1}+C_{2}}{8\gamma
^{2}}}\,.  \label{para}
\end{equation}%
This shows that $\varepsilon _{0}^{-}(q_{x})$ is parabolic at long
wavelengths instead of linear as usually expected for a boson system. This
agrees with the results in Ref. \cite{zhai2}. This parabolic excitation has
a far-reaching consequence: according to Eq. (\ref{cerenkov}), the critical
dragging velocity $v_c$ is zero, very different from the critical velocity
for a BEC moving in a tube. This shows that the critical velocity for a BEC
with SOC is not independent of the reference frame, in stark contrast with a
homogeneous superfluid without SOC. This surprising finding of course has
the root in the fact that the BEC described by the SOC Hamiltonian (\ref{ham}%
) is not invariant under the Galilean transformation \cite{messiah}.

\textit{Rashba and Dresselhaus SOC}:~We have also investigated the superfluidity with the
general form of SOC, which is a mixture of Rashba and Dresselhaus coupling.
Mathematically, this SOC term has the form $\alpha\sigma_{x}p_{y}-\beta
\sigma_{y}p_{x}$. The essential physics is the same: the critical flowing
speed is different from the critical dragging speed, and therefore the
critical velocity depends on the choice of the reference frame. However, the
details do differ when $\alpha \neq \beta $. First, the critical dragging
speed is no longer zero. Secondly, the BEC with negative cross helicity can
also be stable; as a result, there is a different critical speed for either
cross helicity. For the Rashba type ($\alpha=\beta $), the critical speed
for negative cross helicity is always zero.

\textit{Experimental observation}:~Spin-orbit coupled BECs
have been realized recently by three different groups \cite{lin2,fu,pan}
through coupling ultra-cold $^{87}$Rb~atoms with laser fields. The strength
of the SOC in the experiments can be tuned by changing the directions of the
lasers \cite{lin2,fu,pan} or through the fast modulation of the laser
intensities \cite{yongping2}. The interaction between atoms can be adjusted
by varying the confinement potential, the atom number or through the
Feshbach resonance \cite{Chin}. For the scenario in Fig. \ref{soc}(b), one
can use a blue-detuned laser to mimic the impurity  for the measurement of the
critical dragging speed similar to the experiment reported in Ref. \cite{raman}.
For the scenario in Fig. \ref{soc}(a), there are two possible experimental setups for
measuring the critical flowing speed.  In the first one, one generates
a dipole oscillation similar to the experiment in Ref. \cite{pan} but with a blue
detuned laser in the middle of the trap. The second one is more complicated:
At first, one can generate a moving BEC with a gravitomagnetic
trap \cite{motion}. Afterward, one can use Brag spectroscopy \cite%
{brag1,brag2} to measure the excitations of the moving BEC, from which the
superfluidity can be inferred. For the typical atomic density
of $10^{14}\sim10^{15}$ cm$^{-3}$ achievable in current experiments \cite{raman}, the critical
flowing velocity is $0.2\sim0.6$ mm/s, while the critical dragging velocity is still very small, about
$10^{-3}\sim10^{-2}$ mm/s. To further magnify the difference between the two critical
velocities, one can use the Feshbach resonance to tune the $s$-wave scattering
length.

\textit{Perspective}: With BEC, we now have many types of superfluids, which are different from
the conventional spinless and homogeneous superfluid typified by liquid helium.
These new superfluids can be classified into two groups according
to how much they differ from the conventional superfluid.  We call the first group Landau
superfluid, for which Landau's argument (Eq. \ref{landau}) can be used to find its critical
velocity; we call the second group non-Landau superfluid, for which Landau's argument
is either insufficient or invalid for finding its critical velocity. We offer two examples.

The first is a two-component BEC without SOC. As discussed previously,
although this superfluid has two branches of phonon excitations and appears
very differently from the conventional superfluid, its critical velocity can still
be derived from the excitation at the ground state with Landau's argument. Therefore,
this is a Landau superfluid.

\begin{figure}[!h]
\center{\includegraphics[width=8cm]{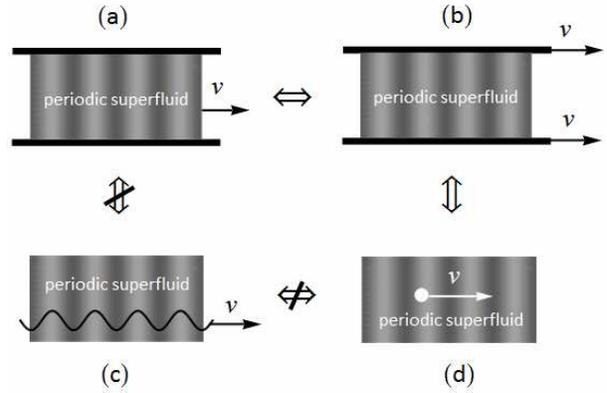}}
\caption{In (a,b,d), the periodic superfluid and the lattice are locked in and move together relative
to the obstacle. In (c), the lattice is gradually accelerated to the speed of $v$.
The reference frame is the lab. The scenario in (c) is not equivalent to the scenarios in (a,b,d). }
\label{periodic}
\end{figure}

The second is a BEC in an optical lattice or supersolid helium. These periodic superfluids
differ from the conventional superfluid by having a periodically modulated density.
We consider first the case when the BEC is locked
with the lattice and they move together as shown in Fig. \ref{periodic}(a,b,d).
Since this system is invariant under the Galilean transformation, the two scenarios
depicted in Fig. \ref{periodic}(a,b) are equivalent and Landau's argument is still
applicable. The caveat is that the critical velocity is always zero no matter how
the elementary excitation of the superfluid looks. The key reason is that the
momentum $\mathbf p$ of the excitation is not well defined due to the presence
of the lattice: two momenta differ by a reciprocal lattice vector are
equivalent. This result has been verified in extensive
computations for Cerenkov radiation in a periodic media \cite{key2}.

The addition of a lattice does put a new twist into the system. One
can gradually accelerate the lattice to a certain velocity and see how the superfluid
changes. There exists a critical velocity of the lattice beyond which the system loses
its superfluidity. This critical velocity can no longer be determined by Landau's
argument as having been discussed in Ref. \cite{wushi}. Therefore, overall a periodic
superfluid is a non-Landau superfluid.

Note that a more detailed version of the above discussion for the superfluidity
of a periodic superfluid can be found in Ref. \cite{wushi}. However, an error was made
in Ref. \cite{wushi}: a non-zero critical velocity for the scenarios shown
in  Fig. \ref{periodic}(a,b,d) was predicted.

The superfluid with SOC studied in this Letter is clearly also a non-Landau superfluid but
for a very different  reason, the lack of invariance under the Galilean transformation.

\acknowledgments
We acknowledge helpful discussion with Hongwei Xiong. Q.Z. and B.W. were
supported by the NBRP of China (2012CB921300) and the NSF of China
(10825417),  and the RFDP of China (20110001110091). C.Z. was
supported by DARPA-YFA (N66001-10-1-4025), ARO
(W911NF-09-1-0248), and NSF-PHY (1104546).

\end{document}